# On Challenges of Cloud Monitoring

William Pourmajidi
Department of Computer Science, Ryerson University
Toronto, Canada
william.pourmajidi@ryerson.ca

John Steinbacher
IBM Canada Lab
Toronto, Canada
jstein@ca.ibm.com

Tony Erwin
IBM Watson and Cloud Platform
Austin, USA
aerwin@us.ibm.com

Andriy Miranskyy
Department of Computer Science, Ryerson University
Toronto, Canada
avm@ryerson.ca

## Abstract

Cloud services are becoming increasingly popular: 60% of information technology spending in 2016 was Cloud-based, and the size of the public Cloud service market will reach $236B by 2020. To ensure reliable operation of the Cloud services, one must monitor their health.

While a number of research challenges in the area of Cloud monitoring have been solved, problems are remaining. This prompted us to highlight three areas, which cause problems to practitioners and require further research. These three areas are as follows: A) defining health states of Cloud systems, B) creating unified monitoring environments, and C) establishing high availability strategies.

In this paper we provide details of these areas and suggest a number of potential solutions to the challenges. We also show that Cloud monitoring presents exciting opportunities for novel research and practice.

## 1 Introduction

While the term "Cloud" has been used in different contexts such as networking and switching, it was Google's CEO Eric Schmidt who used the term in 2006 to refer to a business model that provides various services to consumers over the Internet [42]. With advancements in virtualisation, companies, such as Amazon, Google, IBM, and Microsoft, started to create massive data centres. Cloud providers use these data centres to convert physical computing resources to virtual resources and then provide them to consumers via Internet. Prior to the Cloud environment, application providers had no option but to design and deploy an expensive hardware platform. In addition to the major upfront cost, providers had to plan for an ongoing maintenance cost as well. With a unique set of features such as elasticity, just in time, and pay-per-use, Cloud has become an inevitable platform for computing. All these benefits made Cloud services ubiquitous, leading to increased investment in the Cloud infrastructure: in 2016, companies spent 60% of their information technology budgets on Cloud-based offerings [23]; public Cloud service market will grow to $236B in 2020 with the 22% compound average growth rate [4].

Both, Cloud providers and developers leveraging Cloud offerings, should constantly monitor the health of their products to ensure that their users are satisfied with the service (meeting quality dimensions: availability, reliability, performance, etc.). Thus, monitoring is an essential component of a Cloud platform, and it plays a vital role for Cloud providers and Cloud consumers.

Users who deploy their services on the Cloud require detailed monitoring data and need to keep track of all details of their platform in real-time. In the case of companies with a large number of customers, like Netflix, the metrics that need to be monitored can easily produce more than 10 billion records a day [3] making the data set large enough to be classified as Big Data [27, 25]. The high volume and velocity of generated monitoring data pose various challenges for monitoring systems. Not only the storage of such data is a challenge, but also the data processing portion is computationally expensive [39, 25]. Hence, monitoring large-scale Cloud environments is one of the major challenges of Cloud monitoring [39].

Elasticity, as one of the most favourable features of Cloud computing, is changing the static nature of Cloud platforms to a dynamic environment, where resources are automatically provisioned or decommissioned. Conventional server deployments use a static set of resources and, therefore, a static monitoring tool can provide real-time monitoring for such environments. Many monitoring tools need to be aware of the existence of a resource before they can provide a monitoring service for it. Conventional monitoring tools cannot be used in an elastic environment as the number of resources and their types are changing dynamically [39].

Cloud services are relying on state-of-the-art network facilities that are often more complicated than traditional network deployments. Since network monitoring is an essential practice in any large-scale network, Cloud networks require even more comprehensive monitoring techniques. The majority of existing Cloud monitoring systems focus on computing resources such as CPU and RAM and do not provide a solution for comprehensive network monitoring to capture statistics from various switches, routers, and other network devices. One possible alternative is to use existing network-specific monitoring systems. However, unique features of Cloud networks (such as automatic provisioning) cause accuracy, scalability, and reliability issues for such tools [31].

Cloud computing offers various types of delivery options [24]. Infrastructure as a Service (IaaS) is the most complete form of Cloud delivery, where consumer is provided with full control over the entire life-cycle of virtual machines (VMs). Hence, IaaS requires extensive monitoring systems that cover infrastructure, operating systems, and application metrics. As for the Platform as a Service (PaaS) consumers, they have control over one or more scalable application development and deployment environments and require access to operating metrics of such platform. In contrast, Software as a Service (SaaS) consumers only use applications that are hosted on the Cloud and require very limited monitoring resources [9].

Cloud monitoring systems are designed and implemented by the Cloud providers. Autonomous control over the monitoring systems offers providers with full visibility over their resources. In contrast, Cloud consumers do not have this luxury and are limited to the features that the Cloud providers have offered to them. In other words, the tool that Cloud users use to monitor their resources and their cost is dependent on the Cloud provider. Although, Cloud consumers generally trust Cloud providers, nevertheless, on many occasions a Cloud consumer needs to validate monitoring and billing details given by the Cloud provider. To be objective, this validation requires an independent tool so users can truly verify the accuracy of the data shared by the Cloud provider [30]. Therefore, having an independent solution, which can be used by Cloud providers and consumers at the same time, is crucial. The goal of creating such a solution can be achieved by using external services; such as DataDog [10]; or building your own aggregators using a stack of ElasticSearch [12], Logstash [14], and Kibana [13], (so-called ELK stack [2]); or a stack of InfluxDB [18] and Grafana [21].

Having said that, monitoring of the systems of a Cloud provider may be more important than that of a single user, as the failure of the provider's infrastructure will affect a large number of businesses that use the Cloud services. For example, Amazon Simple Storage Service (S3) is used by more than 200,000 unique domains [34]. On 28th of February 2017, Amazon S3 experienced a massive outage which led to disruptions of many services including Quora, Sailthru, Slack, Giphy and many more [15].

The majority of online distributed systems are deployed on well-known Cloud providers, such as IBM, Amazon, Google, and Microsoft. These providers use traditional reliability solutions, such as logging and 3-way replication [5]. Nevertheless, traditional reliability systems were designed for far less complex information systems and assumed that only a few components require monitoring and only a few components would fail. These assumptions no longer hold and the scale of the monitoring solutions for Cloud systems has to change [5].

## 1.1 Examples of solved challenges

Challenges discussed above have been tackled by researchers. As the Cloud offerings have to scale elastically [39], efforts have been made to build monitoring tools using multi-tier and peer-to-peer architecture, making the tools more resilient to elasticity than conventional monitoring systems [39].

The massive scale of resources on the Cloud and the number of metrics that needs to be monitored can easily exceed billions of records per day [3]. To preserve space required to store logs, Anwar et al. suggest to avoid storing repetitive values, leading to reduction of the size of stored data by up to 80% [3].

Topology and scale of Cloud networks can change dynamically [31]. Traditional network monitoring tools do not support the dynamic nature of Cloud networks. Pongpaibool et al. [31] built fault-tolerance monitoring system based on clustered architecture, improving performance of monitoring applications for Cloud networks.

Cloud computing is delivered through various enabling services of IaaS, PaaS, and SaaS; each of these delivery methods has a different monitoring requirement [9]. Rodrigues et al. [9] focused on these requirements, acknowledging the need for complex monitoring systems to process complicated monitoring scenarios. They also presented an overview of the Cloud monitoring concepts, structure, and solutions.

Implementation of a complete monitoring system requires full access to the components that will be monitored [30]. Nguyen et al. [30] indicated that only Cloud providers have such level of access to Cloud resources. Therefore, most of the Cloud monitoring solutions are built by Cloud providers. In contrast, Cloud consumers require full details of monitoring data and need a way to verify the monitoring details that are provided by the Cloud provider. To address this issue, Nguyen et al. [30] combined role-based monitoring templates with agent-based monitoring and used event processing engine to refine the collected data and to provide a trustworthy and holistic monitoring solution.

Cloud platforms consist of large number of hardware and software components [5, 25], generating large volume of logs and metrics data that exceeds the level that a human can interpret [5, 26]. To address this challenge, Bhattacharyya et al. [5] built metrics anomaly detection system (based on recurrent neural networks) successfully detecting up to 98.3% of anomalies.

## 1.2 Position

As we can see, some challenges were solved; however, additional challenges remain. Based on the authors experience, practitioners experience difficulties with A) defining health states of Cloud systems, B) creating unified monitoring environments, and C) establishing high availability strategies. These challenges have not been fully solved by the research community. Thus, our **position** is that **further research is required in the realm of Cloud monitoring**.

The rest of the paper is structured as follows. Section 2 provides details on the three challenges and highlights some of the potential solutions. Section 3 concludes the paper.

# 2 Challenges and Solutions

In this section. we will discuss issues related to Cloud monitoring that have not been fully solved by the researchers and practitioners. Section 2.1 deals with health states of Cloud environments, Section 2.2 – with unified monitoring environments, and Section 2.3 – with high availability strategies.

Throughout this section, we will be using the term **component** that describes hardware or software component of a Cloud platform, solution, or system. Note that the types of components are numerous and can range from bare-metal computers and network switches, to VMs and containers, to database systems, middleware servers, and front-ends. Moreover, the components may be offered via different service models (e.g., IaaS, PaaS, or SaaS) and vendors (e.g., Amazon, Google, IBM, or Microsoft). While each of these components may have its own unique set of health-related attributes and health states (as well as associated health criteria) [9], the general issues discussed below are applicable to all of them.

## 2.1 Defining health states

In this section, we will discuss issues related to defining health states and establishing mapping between health criteria and various health-related attributes of the components. Typically, the health of a piece of hardware or software component of a Cloud platform or system can be classified using binary states (e.g., "healthy" or "unhealthy"). Once the state classification is selected, Operations personnel (OP) specifies a set of attributes that characterises the health of the component and the thresholds for these attributes that help to classify the state of the component.

**Example 2.1.** *Consider a simple example. For a given VM, OP can set the following criterion: if more than 1GB of space is left on the hard drive where the OS partition resides, then this VM (or, using our terminology, component) is healthy. Once the free space on the partition drops below 1GB, this is a cause for concern, as the space on the hard drive will, eventually, fall to zero and the system will become unresponsive.*

*Let us assume that OP uses a sophisticated monitoring framework similar to DataDog [10], which can alert OP using various communication channels once the state of the VM is changed (e.g., via email, Slack, or pager). Let us also assume that for this VM, OP chose binary health classification ("healthy" or "unhealthy"). Thus, once the space falls below the 1GB threshold, the system becomes "unhealthy", the alarm is issued and delivered to OP. The OP needs to react to the alarm, find the root cause of space reduction, and restore the component back to the "healthy" state.*

*If the OP does not react to the alarm fast enough — the free space on the hard drive will eventually fall to zero and the system will[1] become unresponsive. Once the system becomes unresponsive, the service outage will occur. The repair time in this case may be significant: in the classic data centre, equipment resides on premises (or close to it) and OP can access the hardware directly and perform the repair as needed. Unfortunately, in the Cloud the component resides at a remote location to which, often, OP will not have direct access. Opening a support ticket with the operations team supporting the Cloud hardware will waste more time and prolong the outage; hence the importance of repairing the system before it becomes unresponsive.*

*By Murphy's law, the alarm in our example was raised at 2am. Thus, OP had to be dragged out of bed to investigate and correct the problem. When such event occurs frequently, the productivity of the OP reduces, as productivity of sleep-deprived humans is below par.*

The question then becomes: could we postpone the investigation until normal business hours, say, 9am? That way we do not have to drag anyone out of bed, thus, improving productivity and energy levels of the personnel.

If the hard drive space fills up slowly (say, at a constant rate of 1MB per hour), we could have safely waited another 7 hours. However, if it fills up at a rate of 1GB per hour, then OP would have had to react immediately.

Potential solutions are numerous. One can make classification more nuanced, by introducing ternary states classification (e.g., "green/good", "yellow/warning", "red/bad" that services, such as DataDog [10], have). The "warning" state can be mapped to "wait till 9am" decision while the "red" state can be mapped to "look at it right now" decision.

But how do we differentiate "warning" state from "red" state? For that we need to know if the hard drive will fill up before 9am or not. One can build a model to predict (e.g., using time series forecasting approach [33]), how fast the storage will fill up. However, the model will be based on the assumption that historical behaviour can be used to predict future behaviour, which is not always the case. Say, the system's workload can suddenly intensify (e.g., during a cyber-attack) resulting in a dramatic increase in the number of events that should be logged and stored on the hard drive. Such workload changes are typically rare[2], and prediction models have hard time detecting such changes (as there are not enough occurrences / data associated with rare events).

---

[1] Modern operating systems are resilient and will not become unresponsive in this situation, but remember that this is a toy example.

[2] Note that we need to differentiate rare but repeating events from "black swan events" [35] that rarely repeat themselves and are extremely difficult to predict [35]. For example, consider outage of Amazon S3 service in February of 2017, causing widespread outage for AWS users relying on the S3 service [1]. It was caused by a human error [1] and was almost impossible to predict; however, once known addition checks were put in place to prevent the same event happening in the future [1].

Another issue comes from the fact that OP may be dealing with thousands of components (containers, VMs, switches, etc.). The problem exacerbates when one is using a microservice architecture rather than monolithic architecture, as the number of individual "moving parts" to monitor grows rapidly. Simply storing and rapidly processing logs of the monitoring data becomes a challenge [25]. This can be alleviated by building scalable monitoring infrastructure [39] or outsourcing it to an external scalable service, such as DataDog [10].

In practice, one wants to capture more sophisticated health criteria involving multiple attributes of the component, e.g., if average I/O writes > 1000 wps and HDD space < 1GB and CPU utilisation > 95% then the component is unhealthy. Alas, the task of capturing all possible permutations of the attributes becomes formidable due to combinatorial explosion of permutations of attributes.

However, OP still needs to build proper alarms. It is difficult to build customised alarms to each individual component, when one has thousands[3] of them. Thus, one can start with basic alarm templates loosely tailored to different groups of components under specific workloads (e.g., a set of alarms for Docker containers [11] processing user authentication). Of course, creating these starting templates is a challenge in itself. Once the templates are available, one can then leverage Machine Learning techniques (such as cognitive computing [28] or deep neural networks [5, 17]) to build customised monitoring profiles, tailored for that particular component. The models can be further trained, using reinforcement learning schemes [40], by getting feedback from OP (which can validate alarms emitted by the automatic system, highlighting true, false, and missed ones), when the models issue a false alarm or miss a novel abnormal event.

An issue comes from the fact that such automatic systems require large volumes of data to be trained on; however, we may not have enough data that captures abnormal behaviour, as discussed above. One can try to use an approach for working with imbalanced class, such as under- or over-sampling (see [41] for comparison). This may help in training the model to recognise known rare events but will not help OP to detect new ones (as under- and over-sampling algorithms typically generate data points within the space of the available examples).

Another option is to train a model only on healthy state data, so that it detects outliers (anomalies), i.e., new data points that do not belong to the healthy state [7]. However, the anomaly detection models are less useful to OP, as the model will tell OP that something is wrong, but typically will not be able to identify problematic subset of attributes from the set of all collected attributes (until OP manually examines them).

Another issue comes from the fact that the end-users are typically interested in the health of the overall system rather than individual components. Thus, one needs to compute health state of the overall system, based on the health states of individual components. For example, if a load balancer that processes incoming user requests is down, the health of the remaining components is irrelevant to users, as they cannot access the system. On the other hand, a persistent storage component that processes user requests sub-optimally may be relatively harmless, as long as a user does not notice significant performance degradation.

To compute the health criteria, one may leverage tools from the statistical process control [29], which is encouraged to be used in software engineering by the Capability Maturity Model Integration (CMMI) Levels 4 and 5 [36]. For example, one can formulate the problem of defining the healthy state of the overall system as a tolerance design problem [32] used to compute permissible limits. One can compute permissible limits of the health of every component that would minimise the cost of construction and maintenance while keeping adequate health of the overall system (constrained by the system's Service Level Agreements) using optimisation approaches similar to the ones used in [16, 19].

In addition to the above-mentioned issues, if the monitoring solution fails, then OP will no longer be able to label the state of systems as "healthy" or "unhealthy". To err on the side of caution, resources that are being monitored by a failed monitoring system should be deemed "unhealthy". This assumption is the safest possible assumption (as it is less risky to treat "unknown" as "unhealthy").

## 2.2 Creating unified monitoring environments

One of the major challenges related to Cloud monitoring is that monitoring applications, traditionally, are designed and implemented for a specific discipline [9]. That is, network monitoring applications cannot be used for application monitoring purposes, and Web service monitoring applications cannot be used for server performance monitoring. Due to this issue, both, companies providing Cloud services and companies deploying solution on the Cloud, end up with a handful of Cloud monitoring solutions, each of which acts as an independent tool.

---

[3]This is not uncommon, especially if a solution is implemented using microservice architecture.

Such segregation causes negative consequences. Separate monitoring tools may generate redundant logs and monitoring data. Each monitoring system will have an independent data repository, whose data schemas will be incompatible with each other. Thus, OP has to "hop" between multiple data repositories and dashboards to monitor the systems and diagnose the problems, which is sub-optimal and leads to a waste of resources and to delays in detection and elimination of problems.

One obvious solution is to invest in the creation of a unified monitoring framework. However, both, Cloud services and large solutions deployed on the Cloud, typically require multiple maintenance and operations teams. These teams are often semi-independent, each one having their own cultural[4], legacy, and architectural constraints. This, based on the authors' experience, makes it difficult to migrate to a unified solution; such as a stack of ElasticSearch [12], Logstash [14], and Kibana [13] (ELK stack [2]); or a stack of InfluxDB [18] and Grafana [21]. Moreover, the cost of migration of legacy logging mechanisms and dashboards to a new framework may be prohibitively high.

The task of unification is also complicated by the fact that one may leverage different service providers and service models. For example, an organisation building SaaS offering may be using a combination of IaaS offerings from one vendor and PaaS offerings from another vendor. This makes it even more challenging to create an amalgamated view of all the monitored data for the final SaaS offering.

Cloud platforms consists of several hierarchical layers [37]. The lowest layer is the hardware level and consists of data centre components. Many software-defined layers are implemented on top of this layer [20]. These layers act as the foundation for the application layer which is the closest one to the end-user. While there are clear boundaries among these layers, an issue in a lower layer can easily affect upper layers. One common challenge for cloud providers is to trace an issue and find out the layer (and a component specific to that layer) that is the root cause of a given problem. However, addressing this challenge requires extensive traceability and a holistic view of how layers are inter-related. Note that Cloud providers do report issues that their services experience. However, not every issue is affecting every system that uses this service. Thus, OP that maintains a system using the service often left wondering: is my outage caused by one of the issues reported by my service provider or is it the problem with my own system?

An additional challenge is that there is no worldwide standard for monitoring architectures, even though a number of architectures and interfaces have been proposed [6, 8, 22]. Therefore, defining and enforcing a single standard may raise additional difficulties.

## 2.3 Establishing high availability failover strategies

Defining failover and high availability (HA) for large distributed systems is a non-trivial task, as the systems generally consist of several clusters, database systems, availability zones, complex networks, and a combination of monolith and microservice architectures. Implementing HA policies and strategies requires extensive monitoring of resources. Monitoring solutions that will be used in HA architecture, need to provide a holistic health state of the main site as well as all the backup sites that will be used for failovers. Such a broad coverage tremendously increases the number of elements that a monitoring solution should cover. While the implemented monitoring solution should handle the collection of such massive number of metrics and elements in real-time, it should also remain as responsive as possible, so that the HA components can make decisions based on the latest and the most accurate state of each element. Finding a balance between the minimum number of elements that need to be monitored and the performance of monitoring system requires further research.

Failovers are computationally and commercially expensive and should be only used when necessary. Therefore, deciding whether to failover or keep using the same resources becomes an important topic that requires further exploration. It is closely related to the discussion on the health state of the system (discussed in Section 2.1), as the health state is used by HA solutions to decide when to initiate a failover. Once it is determined that a failover is required, selecting a failover strategy becomes critical. In the simplest case, we can failover the whole system, say, by redirecting traffic to a different instance of the system residing in a different geographical zone. However, this may dramatically increase load on the instance to which we redirected traffic, causing its performance to degrade (and potentially an outage due to overload).

---

[4]In a team- or corporate-culture sense.

Probabilistically, rarely all components of a large system will malfunction simultaneously. Thus, potentially one can often failover just a portion of the system rather than the whole system (which may be faster and less expensive than failing over the whole system). Therefore, selecting the portion of the platform that requires failover becomes the most important task. It will be critical to understand[5] what has gone wrong and which components are affected, so a logical failover plan is developed and executed. To the best of our knowledge, no general framework to address this task exists at the time of writing.

When it comes to designing and implementing HA for services (for both monolith and microservice architecture), a major challenge arises if the transactions[6] are distributed among multiple nodes. If one node fails, a usual failover approach may bring back the service, but the state of the transaction may become unknown and/or the data related to that transaction may become inconsistent. A primary goal for failover practices is to retrieve the service as soon as possible, and, at times, this may favour service over data consistency. For example, consider a VM that hosts a service which relies on in-memory database. If this service experiences a deadlock situation, restarting the VM would bring back the service at the cost of losing the data and state which were saved in memory.

The monolith architecture has a somewhat relaxed approach towards the number of business tasks that are handled by each service. It is very common to see a server-side monolithic component that handles all the business tasks and has internal logic to ensure atomicity of processed tasks. In contrast, in the case of microservices, ideally, we would like to keep transaction limited to a particular microservice, which makes failover and transaction rollback straightforward. However, architecturally it is not always possible. In such a case, a potential solution would be to create a monitoring service that satisfies Atomicity, Consistency, Isolation, Durability (ACID) principles and will not allow failovers that would result in data inconsistency. Following a series of predefined rules, in case of a microservice failure, the monitoring system will make a logical decision about failing over one, a group, or all of the microservices based on transactional boundaries.

In addition to the above challenges, each component of the system has a different degree of importance in regard to HA and failover policies. For example, if a VM running a service fails, an acceptable failover policy would be to launch a new VM on another server using the same image. The same logic applies to bare-metal servers. A faulty bare-metal server can be easily replaced by another bare-metal server, or in some cases a VM, in a different server farm. In contrast, when the main load balancer is down, or if the default gateway of a cluster is down, irrespective of the state of each individual component, the failover policy can no longer use the same resources in the cluster as none of them are accessible.

Another potential solution is to use application-specific HA structure. The architects of the application create a tiered structure of components. Then, a set of failover polices will be defined for each tier. At the time of failure, the monitoring application will identify the tier and take necessary actions that are defined within the application-specific HA rules. By adding a context-aware, software-defined monitoring solution, we can define a smart failover plan that: 1) considers the device and its tier before calling a failover action, 2) can decide whether to failover the entire application or just one microservice, and 3) can prevent data loss in case of failure of a microservice (by applying ACID-like features to the monitoring system, as discussed above).

Finally, changes to the software may cause regression of functionality. Twitter experienced a major outage due to such an event on 19th of January 2016 [38]. If the breaking software change has been delivered to both primary and failover systems, then failover will not help and the only remaining course of action is to roll back the change[7] and restart the system. Thus, one should be careful about how the code changes are being rolled out to production systems. Ideally, rollback can be integrated into the HA strategies. However, it is unclear if this can be automated for any type of regression-inducing change. For example, changes to source code of a script can be rolled back automatically with ease. However, changes to a data schema and the data itself may require manual intervention.

---

[5] Note that this will require access to the health data for all the components of the system, which may be challenging, as we discussed in Section 2.2.

[6] Which we informally define as the smallest unit of atomicity that can represent a business goal.

[7] Or fix the regression, which may be riskier as developers will not have a lot of time to test the change.

# 3 Summary

In the past, application providers had to design and deploy their products on a costly hardware platform. The situation has changed with introduction of Cloud services that led to reduction of upfront costs, improved elasticity, etc. These benefits led to dramatic increase of popularity of Cloud services (60% of information technology spending in 2016 was Cloud-based [23]). To guarantee reliable operation of the Cloud offerings, one must monitor their health.

In this paper, we highlight some of the challenges posed by the task of monitoring Cloud solutions, platforms, and systems. While some challenges have already been solved, our position is that further research is required in the realm of Cloud monitoring.

In this work, we focused on three areas: A) defining health states of Cloud systems, B) creating unified monitoring environments, and C) establishing high availability strategies. We highlight some of the unsolved issues in these areas and suggest research tools that may be used to solve these issues in the future. We also show that these issues are interconnected: to make HA decisions (area C), one needs to understand health states (area A) of all components of a Cloud platform, solution, or system (area B).

Thus, Cloud monitoring is a fertile area for novel research and practice. Solving the above challenges will simplify integration of Cloud monitoring into the maintenance and operations phases of the software development life cycle, while reducing the risk of outages, lessening maintenance cost, and decreasing the load on human and computer resources.

## Acknowledgements

The authors would like to thank IBM Centre for Advanced Studies for their generous support and the reviewers for their insightful comments.

This research is funded in part by IBM CAS Project No. 1046 and NSERC Discovery Grant No. RGPIN-2015-06075.